\documentclass[useAMS,usenatbib]{mn2e}
\usepackage{graphicx}
\usepackage{epstopdf}
\usepackage{mathrsfs}
\usepackage{amssymb}
\usepackage[title]{appendix}
\usepackage{amsmath}
\usepackage{color}

\newcommand{\Mpc}{{\ensuremath{\rm Mpc}}}

\newcommand{\pri}{\prime}
\newcommand{\tr}{\rm tr}

\newcommand{\vu}{\mathbf u}

\newcommand{\bvu}{\bar{\vu}}
\newcommand{\bA}{\bar {A}}
\newcommand{\bs}{\bar {s}}

\newcommand{\vv}{\mathbf v}

\newcommand{\vx}{\mathbf x}

\newcommand{\vT}{\mathbf T}
\newcommand{\vI}{\mathbf I}

\newcommand{\vq}{\mathbf q}

\newcommand{\vpsi}{\mathbf \Psi}

\newcommand{\mH}{\mathcal H}
\newcommand{\Om}{\Omega_m}

\begin{document}


\title[Nonlinear Evolution of Cosmic Web]{On the Nonlinear Evolution of Cosmic Web:
Lagrangian Dynamics Revisited}

\author[X. Wang et al.]{Xin Wang$^{\dagger*}$, Alex Szalay$^{\dagger}$
\vspace*{4pt}\\
$^1$ Department of Physics \& Astronomy, Johns Hopkins University, 
Baltimore, MD, US, 21218\\
$^*$ wangxin$@$pha.jhu.edu
}

\date{Accepted . Received ;}

\pagerange{\pageref{firstpage}--\pageref{lastpage}} 

\maketitle

\label{firstpage}

\begin{abstract}
We investigate the nonlinear evolution of cosmic morphologies of the large-scale 
structure by examining the Lagrangian dynamics of various tensors of a cosmic 
fluid element, including the velocity gradient tensor, the Hessian matrix of the 
gravitational potential as well as the deformation tensor. 
Instead of the eigenvalue representation, the first two tensors, which associate 
with the `kinematic' and `dynamical' cosmic web classification algorithm respectively, 
are studied in a more convenient parameter space.  
These parameters are defined as the rotational invariant coefficients of the 
characteristic equation of the tensor. In the nonlinear local model (NLM) where 
the magnetic part of Weyl tensor vanishes, these invariants are fully capable 
of characterizing the dynamics. Unlike the Zel'dovich approximation (ZA), where 
various morphologies do not change before approaching a one-dimensional singularity, 
the sheets in NLM are unstable for both overdense and underdense perturbations. 
While it has long been known that the coupling between tidal tensor and velocity 
shear would cause a filamentary final configuration of a collapsing region, 
we show that the underdense perturbation are more subtle, as the balance between 
the shear rate (tidal force) and the divergence (density) could lead to different 
morphologies. Interestingly, this instability also sets the basis for understanding 
some distinctions of the cosmic web identified dynamically and kinematically. 
We show that the sheets with negative density perturbation in the potential based 
algorithm would turn to filaments faster than in the kinematic method, which could 
explain the distorted dynamical filamentary structure observed in the simulation. 
\end{abstract}


\section{Introduction}
As a result of anisotropic gravitational instability, the matter distribution of the
Universe at large scale exhibits an intrinsic pattern, known as the cosmic web 
\citep{BKP96}. 
This particular structure, characterized by a network of filaments, sheets 
and empty voids, has repeatedly been revealed by various observations, 
including the large scale galaxies distribution from galaxies surveys
\citep{GT78,dL86,GH89,SLO96,C03,T04,H05} and the dark matter map inferred 
from weak lensing survey \citep{MR07}. 
In this regard, the presence of this structure in both numerical simulations and
analytical models, e.g. the Zel'dovich approximation \citep[ZA,][]{ZA70}, highlights 
our achievement of understanding the process of structure formation.

Besides its significance in the theory of large-scale structure, the cosmic web 
also serves as an environment for small scale structures like halos and galaxies. 
Various observations suggest that many galaxy properties vary with this environment
systematically \citep{D80,KW04,BE05}. 
Meanwhile, high-resolution numerical simulations also show clear correlations between
halo properties, like concentration and spin, with the local environment 
\citep{LK99,ST04,ACG05,WZ05,BE07,MD07,WCW07,HP07a,HC07b}. 
Hence, it is essential to define properly and better understand this 
morphological environment and its time evolution.

\begin{figure*}
\begin{center}
\includegraphics[width=1.\textwidth]{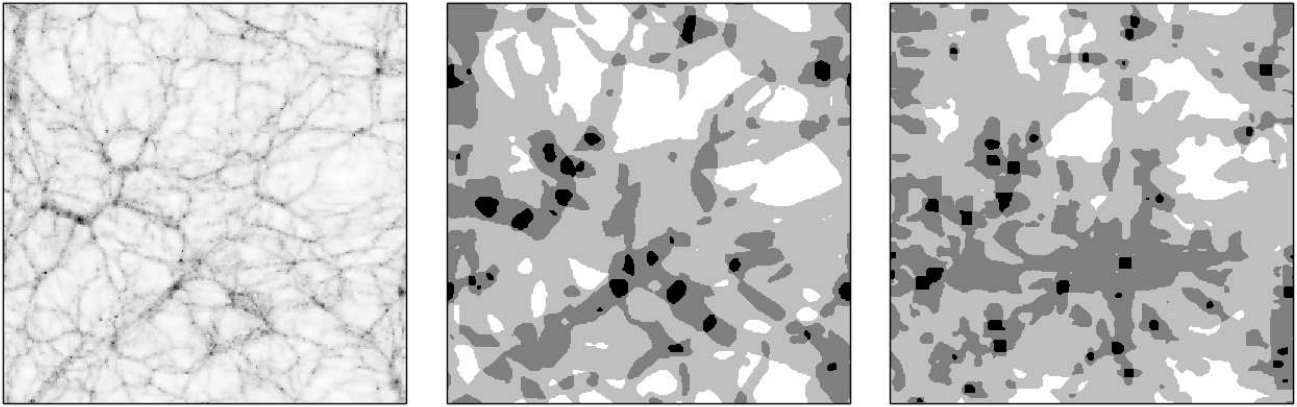}
\end{center}
\caption{ \label{fig:T_V_sim_snapshot}  
Comparison between kinematic ({\it middle panel}) and dynamical ({\it right panel}) classification for one snapshot of N-body simulation with the density distribution 
shown in the first panel. 
Both velocity and gravitational potential fields have been smoothed by a Gaussian 
filter with characteristic scale $R=1\Mpc/h$ before applying the algorithm.
The black, grey and silver regions illustrate knots, filaments and sheets respectively, 
while white shows the voids. 
Unlike some other works, we assume eigenvalue threshold $\lambda_{th}=0$ for both algorithms.
Therefore, different morphologies identified here are not `optimized' to 
match the visual impression. Furthermore, this snapshot is selected in particular to 
highlight the significant differences these two methods could produce. 
The simulation has a box size of $100~\Mpc/h$, with $256^3$ particles. Both density and
velocity field was estimated with Delaunay tessellation \citep{BvdW96,SvdW00,PSvdW03}. 
}
\end{figure*}

Although this web structure originates from anisotropic initial random field
\citep{Dor70} and is qualitatively well described by simple analytical model like
ZA, its evolution, however, is highly nonlinear at later epoch.
Consequently, many efforts have been concentrated on developing algorithms of 
morphology classification for numerically simulated data. 
At least two categories of such algorithms exist in literatures, 
one is geometrical method, which tries to establish a mathematical description 
based on the point samples of galaxies/halos or dark matter particles in simulations
\citep{LK99,NCD06,AJ07,SP08}.
The other `dynamical/kinematic' approach then considers the movement of a test 
particle in the inhomogeneous gravitational potential, which could either be 
described by the Hessian matrix of gravitational potential \citep{HP07a,FR09} or 
the velocity gradient tensor \citep{HM12}. 
Various morphologies can then be identified by the number of eigenvalues greater
than some threshold value.

In Zel'dovich's structure formation theory \citep{ZA70}, these two tensors are simply
proportional to each other, and therefore provide identical morphology classification.
However, once the Universe enters into the nonlinear regime, differences start
to emerge. 
With a similar level of smoothing and comparable eigenvalue thresholds, \cite{HM12} 
showed that the velocity based algorithm provides very different morphologies than 
the gravitational potential based technique and seems to resolve smaller structures 
(also in Figure. \ref{fig:T_V_sim_snapshot} for zero eigenvalue threshold).
However, the reason of such dissimilarity is not clear, and theoretical investigation
is highly demanded.
Furthermore, instead of the null eigenvalue threshold first utilized by \cite{HP07a}, 
subsequent studies in general assume a nonzero threshold which is usually artificially 
tuned to provide the best visual impression. 
For gravitational potential based cosmic web, \cite{FR09} suggested that the 
threshold value of properly normalized tensor should be around unity based on 
the argument of spherical collapse.
However, the justification requires more detailed study on the classification schemes 
and cosmic evolution of both tensors.

Compared with geometrical methods, where algorithmic procedures usually prevent them
from further analytical understandings, one advantage of considering the dynamical 
variables is the potential to quantitatively investigate the cosmic web evolution, 
which however, has not been fully appreciated yet. 
One of the obstacles is the less-convenient eigenvalue representation of the algorithm. 
On the other hand, it is equivalent to study the eigenvalues and the rotational 
invariant coefficients of the characteristic equation of the tensor \citep{WS14}. 
As shown by \cite{WS14}, the most important advantage of this parameter space is to 
avoid the complex domain after the tensor become non-symmetric. 
However, even for the purpose of this paper, where only the symmetric part is of 
interest, it will still be convenient to work in this invariant space.
Especially, when the trace of the tensor does not change the sign, a two-dimensional 
subspace would suffice to present morphological evolution.

\begin{figure*}
\begin{center}
\includegraphics[width=1\textwidth]{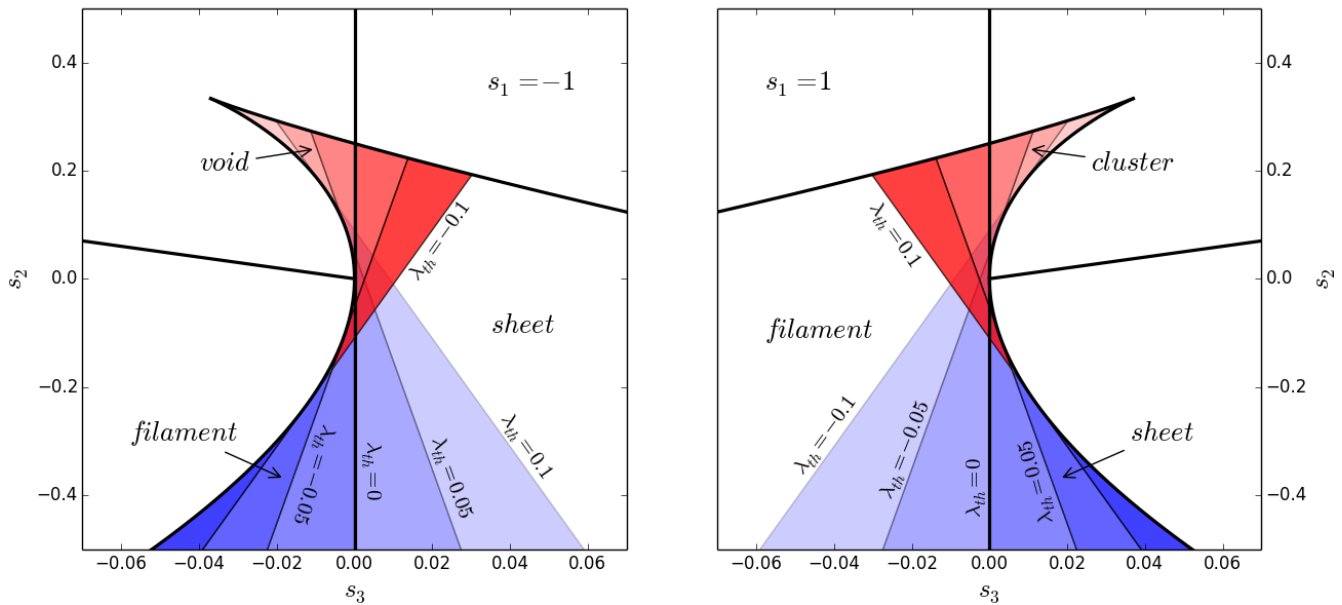}
\end{center}
\caption{\label{fig:inv_space_th} Illustration of cosmic web classification in the 
invariants space with various eigenvalue threshold $\lambda_{th}$. 
We only display two dimensional $s_3-s_2$ planes with fixed $s_1=-1$ ({\it left}) 
and $s_1=1$ ({\it right}). 
Regions with different transparencies highlight the consequences of varying 
eigenvalue threshold, assuming $s_1$ doesn't change the sign. 
Since the boundary conditions depend on the value of $s_1$, we renormalize 
$s^{th}_i$ ($i= 2$ or $3$) by $(s^{th}_1)^i$ after the transformation equation
(\ref{eqn:s_redfine_th}) so that they remain the same. 
For more details of the invariants classification including rotational categories, 
please see Figure (1) in \citet{WS14}.}
\end{figure*}

The main purpose of this paper is therefore to investigate theoretically in this invariant
space, the nonlinear evolution of the velocity gradient tensor and the Hessian matrix 
of gravitational potential. Furthermore, to compare with geometrical 
algorithm, we will examine the evolution of the deformation tensor as well.
To this end, we adopt the Lagrangian approach to track the dynamical evolution of 
relevant variables of a fluid element, including the density, the velocity gradient 
and the tidal tensor.
However in Newtonian theory (NT), the evolution equation of the tidal tensor is missing. 
An alternative approach, as discussed by \cite{BJ94} (hereafter denotes as `BJ94'), 
instead starts from the full general relativistic description, 
where the Lagrangian evolution equations and constraints of gravitational fields are 
well-known. Specifically, the evolution equation of the counterpart of the tidal tensor, 
i.e. the electric part of Weyl tensor, can be derived from the Bianchi identities.
However, the treatment of the magnetic part of this tensor in NT is unclear and 
therefore triggered many discussions \citep{BH94,ED97}.
Nevertheless, in the current paper, we will simply adopt the approach same as 
\cite{BJ94}, assuming the vanishing magnetic part of Weyl tensor, which produces 
a set of self-consistent \citep{LDE95} closed ordinary differential equations.

In this model, the interaction of tensor perturbations between neighboring 
fluid elements is neglected, therefore also known as the `silent universe' model.
Since NT is intrinsically nonlocal as the potential is determined 
by the matter distribution everywhere via Poisson equation, 
a non-general relativistic theory with an extra time evolution equation of the 
tidal tensor should be regarded as some extension of NT, a closer approximation 
to the general relativity \citep{ED97}.
Practically, assuming a vanishing magnetic part of Weyl tensor would dramatically 
simplify the formalism \citep{MPS93,MPS94,BMP95,LDE95}. On the other hand,
one obvious shortage of the Lagrangian approach is the existence 
of the singularity at the shell-crossing. It sets the validity range of the 
method much earlier than the formation of virialized objects. 
Fortunately, for our purpose, it is equally, if not more, important to study the 
underdense perturbations as the visual impression of the cosmic morphologies is 
highly weighted by the lower dense regions for their greater volume filling 
factors.

The paper is organized as follows. In section 2, we first briefly revisit two types 
of cosmic classification algorithms and then introduce the definition of rotational 
invariants as well as the geometrical quantities related to the deformation tensor. 
In section 3, we discuss Lagrangian dynamical evolution models. 
We first show the analytical results of invariants evolution in Zel'dovich approximation 
in section 3.1 and then review BJ94's nonlinear local evolution model in section 3.2 
before deriving the basic equation for obtaining the deformation tensor. 
We present our result in section 4 by first comparing the Zel'dovich approximation 
and the nonlinear local model. We then discuss the differences between velocity 
gradient tensor and Hessian matrix of gravitational potential in section 4.2. 
After discussing the eigenvalue threshold, we finally conclude in Section 5.

\section{Cosmic Web Classification}

We are interested in one particular category of cosmic web classification 
algorithm that considers the movement of a test particle in the anisotropic 
gravitational field. In an appropriate frame, this could either be described by 
the Hessian matrix of gravitational potential or the velocity gradient tensor. 
While the first characterizes the acceleration of the particle caused by the gravity, 
the latter describes the velocity changes. 
Both of them relate to the trajectories of the particle around given point by the 
time integral. 
Given the matrices, rotational invariants provide a more convenient parameter space for
studying the sign of eigenvalues. For nonzero thresholds, however, the transformation 
of the invariants would be necessary, or equivalently the boundary condition among 
various morphologies need to be modified. 
To compare with other geometrical classification approaches, it is also valuable to 
examine the deformation tensor and derived scalars, e.g. the ellipticity and 
prolaticity. These scalars, usually well-defined in the linear region, require 
appropriate revision in the nonlinear regime.

\subsection{Dynamical and Kinematic Approaches} 
As initiated  by \cite{HP07a} and followed by \cite{FR09}, the dynamical approach
considers the linearized equation of motion of a test particle near given position
$\bar{\vx}$
\begin{eqnarray}
\frac{d^2 x_i}{dt^2} = - \Phi_{i}^{~j} (\bar{\vx}) \left(\vx_j - \bar{\vx}_j \right), 
\end{eqnarray} 
where $\vx$ is the comoving free-falling coordinate, $t$ is some time variable,  
$\Phi_{ij} =  \partial_i \partial_j \Phi$ is the Hessian matrix of
peculiar gravitational potential $\Phi$, and the zeroth-order term 
$\partial_i \Phi$ disappears in this frame. 
Therefore, the linear dynamics near location $\bar{\vx}$ is fully characterized by 
tensor $\Phi_{ij}$, or as it's real and symmetric, three eigenvalues of $\Phi_{ij}$.
Then various morphologies could be classified by counting the number of eigenvalues 
greater than some threshold value. 
As motivated by this method, \cite{HM12} proposed a similar 
classification scheme based on the kinematic movement of the particle, also known as 
V-web, which is equivalent to considering the particle trajectory near $\bar{\vx}$ as 
\begin{eqnarray}
\frac{d x_i}{d\tau} = A_{i}^{~j} (\bar{\vx}) \left (\vx_j - \bar{\vx}_j \right)
\end{eqnarray}
where $A_{ij} = \partial_i v_j$. Unlike the potential Hessian matrix $\Phi_{ij}$, in 
principle $A_{ij}$ could be rotational and therefore non-symmetric \citep{WS14}. 
However, in this paper we will only concentrate on the symmetric part of $A_{ij}$, 
assuming $A_{ij} = A_{ji}$.

Both kinematic and dynamical approaches try to identify various cosmic structure 
by examining the tentative movement of the test particle. 
Intuitively, one would expect the deviation between these two approaches after 
entering into the nonlinear regime as the acceleration/deceleration of the particle 
towards a certain direction is not necessarily the same as the velocity. 
In Figure. (\ref{fig:T_V_sim_snapshot}), we compare these two algorithms in the
numerical simulation. 
Before performing the classification, both velocity and potential fields have been 
smoothed by a Gaussian filter with the characteristic length $R=1\Mpc/h$.  
Unlike other works, here we assume the zero eigenvalue threshold for both tensors
so that the cosmic web structure highlighted here would not be `optimized' visually.
From the figure, it is clear that these two methods produce very different 
classifications in details.
Actually, we select this snapshot in particular to underline how significant 
these two approaches could differ.
However, as will be shown in section 4, at least for filaments, the dissimilarity 
could be alleviated by restricting the trace of the tensor.

\begin{figure*}
\begin{center}
\includegraphics[width=1.\textwidth]{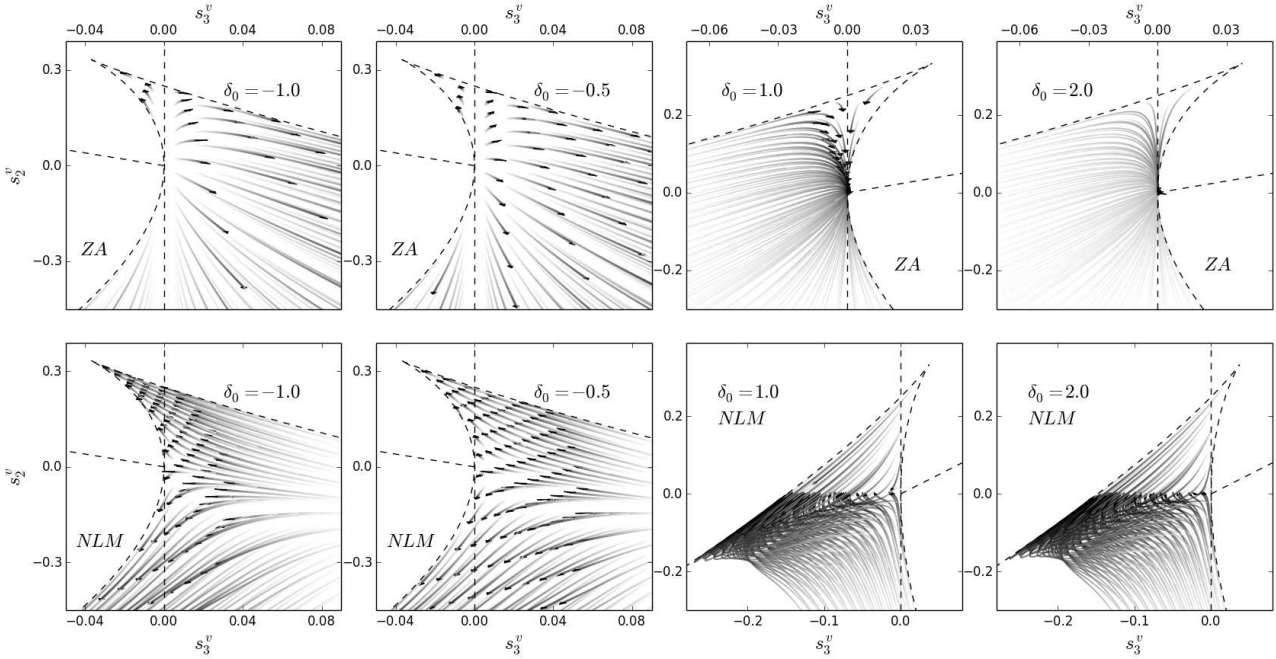}
\end{center}
\caption{ \label{fig:inv_traj_za_BJ94_comp} Comparison between Zel'dovich approximation 
({\it upper panels}) and nonlinear local model ({\rm lower panels}) in the velocity 
invariant space $s^{(v)}_i$.
From left to right, we assume the linear density perturbation $\delta_0=-1,~ -0.5, ~1$ 
and $2$, where $\delta_0 = \delta(a=1)$. 
To better present the result, we have normalized invariants $s^{(v)}_i$ such that 
${\rm tr}[A_{ij}]= \pm 1$. For underdense perturbations all trajectories are plotted 
from the initial epoch $a_0 = 10^{-3}$ to the present $a = 1$, while we stop at the
first singularity for $\delta_0>0$.  
}
\end{figure*}

\subsection{Classification with Rotational Invariants}
In practice, both approaches concern the number of eigenvalues of the tensor 
above a certain threshold $\lambda_{th}$. As proposed by \cite{WS14}, 
assuming $\lambda_{th}=0$, this problem could also be reformulated as 
considering the number of positive/negative solutions of characteristic
equation $\det[\vT-\lambda \vI] = 0 $ of tensor $\vT$, i.e. 
\begin{eqnarray}
\label{eqn:chareq}
 \lambda^3 + s_1 \lambda^2 + s_2 \lambda + s_3=0,
\end{eqnarray}
where we assume $T_{ij} = A_{ij}$ for kinematic method and $-\Phi_{ij}$ for 
dynamical approach. The rotational invariant coefficients $s_1, s_2, s_3$ are 
defined as \citep{CPC90,WS14}
\begin{eqnarray}
\label{eqn:inv_def}
s_1 &=& -\tr[\vT]= -T_{i}^{~i} = -\sum_i \lambda_i, \nonumber \\
s_2&=& \frac{1}{2} \left ( s_1^2 - \tr[\vT^2] \right ) = 
 \frac{1}{2} (s_1^2 - T_{ij}T^{ji}) = \sum_{i\ne j}\lambda_i \lambda_j \nonumber \\
s_3&=& - \det[\vT] = \frac{1}{3} \left (-s_1^3 + 3s_1 s_2 - \tr[\vT^3] \right) 
\nonumber \\ 
&=& \frac{1}{3} \left( -s_1^3  + 3s_1 s_2 - T_{i}^{~j}T_{j}^{~k}T_{k}^{~i} \right)
= - \prod_i \lambda_i.
\end{eqnarray}
Here we have already neglected the anti-symmetric contribution of tensor $T_{ij}$, so 
that $T_{ij}=T_{ji}$. In the last equality of the definition, we have already expressed 
them in terms of real eigenvalues $\lambda_i$. 
In the following, we will denote $s^v$ as invariants constructed from $A_{ij}$ and 
$s^{\phi}$ as from $-\Phi_{ij}$. 
Then the conditions of having certain number of positive/negative eigenvalues, 
is mapped into various regions in the invariants space. 
In Figure. (\ref{fig:inv_space_th}), we briefly illustrate these classifications in thick 
solid lines for both negative ({\it left panel}) and positive $s_1$ ({\it right panel}). 
In both cases, the triangular-like region with real eigenvalues is enclosed by the 
solution of equation \citep{CPC90,WS14}
\begin{eqnarray}
\label{eqn:solution_S1}
 27 s_3^2 + ( 4 s_1^3 - 18 s_1 s_2) s_3 + (4 s_2 ^3 - s_1^2 s_2^2) = 0, 
\end{eqnarray}
with the assumptions  $s_2 \le s_1^2/3$. Within this region, different morphologies 
are basically determined by the sign of invariants. 
For $s_1<0$, region with $s_2>0$ and $s_3<0$ denotes the void, while condition 
$s_2 <0,~s_3<0$ corresponds to the filament, and $s_3 >0$ is cosmic sheet. 
Similarly for $s_1>0$, condition $s_2>0,~ s_3>0 $ defines the cluster, 
$s_2<0,~s_3>0$ corresponds to the sheet, and $s_3<0$ denotes the filament. 
Furthermore, it is worth noticing that the classification would remain the same if 
the invariants are rescaled as
\begin{eqnarray}
\label{eqn:si_normalize}
\tilde{s}_i=s_i/(c)^i, \qquad where ~  c>0
\end{eqnarray} 
by any positive constant $c$, since the normalization of $T_{ij}$ with $c$ would 
not change the sign of its eigenvalues. 
Please see \cite{WS14} for more details of the invariants-based cosmic web classification.

For a nonzero eigenvalue threshold $\lambda_{th}$, however, one could define the 
eigenvalue $\lambda^{\pri}=\lambda-\lambda_{th}$, and rewrite the characteristic 
equation (\ref{eqn:chareq}) as function of $\lambda^{\pri}$, 
\begin{eqnarray}
\label{eqn:chareq_th}
(\lambda^{\pri} + \lambda_{th})^3 + s_1 (\lambda^{\pri} + \lambda_{th})^2 + 
s_2 (\lambda^{\pri} + \lambda_{th}) + s_3 = 0, 
\end{eqnarray}
and then discuss the sign of variable $\lambda^{\pri}$. Expanding equation 
(\ref{eqn:chareq_th}), it is equivalent to define a new set of invariants $s^{th}$
as \citep{WS14}, 
\begin{eqnarray}
\label{eqn:s_redfine_th}
s^{th}_1 &=& s_1 + 3 \lambda_{th}\nonumber\\
s^{th}_2 &=& s_2 + 2 \lambda_{th} s_1 + 3 \lambda_{th}^2 \nonumber\\
s^{th}_3 &=& s_3 + \lambda_{th} s_2 + \lambda_{th}^2 s_1 + \lambda_{th}^3 .
\end{eqnarray}
Then all classification conditions of the invariants remain the same. Alternatively, 
with the help of equation (\ref{eqn:s_redfine_th}), one could also express all 
conditions back into the original invariants space. 
Since any linear combination of real eigenvalues with a real threshold 
$\lambda_{th}$ remains real, the boundary condition separating real and 
complex solutions does not need to vary in the transformed three-dimensional 
invariant space $s^{th}_i, ~i=\{1,2,3\}$. 
However, for illustrative purpose, we are also interested in the two-dimensional
$s^{th}_3-s^{th}_2$ subspace as displayed in Figure (\ref{fig:inv_space_th}), 
where the real-image separation surface would project onto $s^{th}_1=const$ plane 
differently. But if the sign of $s^{th}_1$ remains the same, 
one could simply rescale the invariants so that $s^{th}_1$ is unchanged 
(e.g. $s^{th}_1=\pm 1$), and the same for the boundary in the $s^{th}_3-s^{th}_2$ plane.
Then various morphologies within this region would be classified based on the sign
of $s_3^{th}$ and its intersection with one of the solution in equation  (\ref{eqn:solution_S1}).

In Figure (\ref{fig:inv_space_th}), we also illustrate the effect of varying the 
threshold $\lambda_{th}$ in the same two-dimensional invariants subspace, assuming 
$s^{th}_1$ does not change the sign. 
The thin solid lines correspond to the transformed equation $s^{th}_3=0$, 
and various morphological regions with different $\lambda_{th}$ are then enclosed
by these lines and thick solid boundaries. For example, void or cluster are still
those triangular regions, but with the original vertical condition $s_3=0$ changing
accordingly. In each panel, we also set the transparency level to each threshold 
value for different colored (morphological) regions consistently.
As could be seen, a negative $\lambda_{th}$ will enlarge the region classified as 
void, and shrink the region of cluster and oppositely for positive $\lambda_{th}$. 
For sheet and filament, however, they will be affected a little more complicated. 
As sheets with positive $s_1$ become more abundant with negative $\lambda_{th}$, 
regions tagged as sheet with negative $s_1$ at $\lambda_{th}=0$ might become 
voids, and some filamentary regions will then be identified as sheet. 
Meanwhile, filamentary regions with $s_1<0$ will be reduced, and some filaments 
with $s_1>0$ at $\lambda_{th}=0$ will be classified as sheets. On the 
other hand, a few cluster regions will turn to this type; 
and positive threshold would behave oppositely.

\subsection{Deformation Tensor}
Besides the classification based on $\Phi_{ij}$ and $A_{ij}$, it will also be helpful
to examine the deformation tensor as well, as it characterizes the geometric
deformation of a fluid element during the evolution. The tensor is defined as 
\begin{eqnarray}
\Psi_{ij} = \frac{\partial \Psi_i}{\partial q_j} = J_{ij} - I_{ij}. 
\end{eqnarray}
Here the displacement vector $\Psi_i = x_i - q_j$, where $x_i$ is the Eulerian position
and $q_i$ is initial Lagrangian coordinate. 
$J_{ij} = \partial x_i/\partial q_j$ is the Jacobian matrix between $q_i$ and $x_i$, 
and $I_{ij}$ is the identity matrix.
Unlike classification algorithms introduced previously, tensor $\Psi_{ij}$
characterizes the changes in shape and size of the fluid element.
Hence, it would be convenient to define the geometric quantities like the ellipticity 
and prolaticity from $\Psi_{ij}$. For Zel'dovich approximation, where all eigenvalues 
of $\Psi_{ij}$ grow linearly, they are defined as \citep{BKP96}
\begin{eqnarray}
\label{eqn:ep_lambda}
e^{\lambda}=\frac{\lambda^{\psi}_3-\lambda^{\psi}_1}{|\lambda^{\psi}_1+
         \lambda^{\psi}_2+\lambda^{\psi}_3|}, \qquad
p^{\lambda}=\frac{\lambda^{\psi}_1+\lambda^{\psi}_3-2\lambda^{\psi}_2}
          {|\lambda^{\psi}_1+\lambda^{\psi}_2+\lambda^{\psi}_3|}, 
\end{eqnarray}
assuming $\lambda^{\psi}_3\ge\lambda^{\psi}_2\ge\lambda^{\psi}_1$, where 
$\lambda^{\psi}_i$ are eigenvalues of deformation tensor $\Psi_{ij}$.
However, for nonlinear evolution, it is possible that the denominator 
$\sum_i\lambda^{\psi}_i $ changes the sign during the cosmic evolution 
\footnote{Even for ZA-like model $\nabla_q \cdot \vpsi \propto \delta$, where $\delta$ 
is nonlinear, it is possible that an underdense perturbation $\delta<0$ would 
eventually collapse \citep{BJ94}. }. 
Therefore, $e^{\lambda}$ and $p^{\lambda}$ could be singular even before the 
shell-crossing. 
Therefore, in this paper, we recover the full definition of the nonlinear ellipticity 
and prolaticity by substituting the denominator with its nonlinear counterpart 
$|J-1| = |\delta| $
\begin{eqnarray}
\label{eqn:ep_J}
e^J=\frac{\lambda^{\psi}_3-\lambda^{\psi}_1}{|J-1|}, \qquad
p^J=\frac{\lambda^{\psi}_1+\lambda^{\psi}_3-2\lambda^{\psi}_2}{|J-1|}, 
\end{eqnarray}
where $J$ is the determinant of the Jacobian $J_{ij}$.
For $|\delta| \ll 1$, these two equations are equivalent, and $|J-1| = \prod_i \lambda^{\psi}_i$ would keep the sign before the shell-crossing. 
One notices that the boundary condition $e^J\ge 0$ and $ -e^J \le p^J \le e^J$ 
are still valid.
In the following, we will omit superscripts and simply denote $e$ and $p$ as 
defined in equation (\ref{eqn:ep_J}).

\begin{figure*}
\begin{center}
\includegraphics[width=1\textwidth]{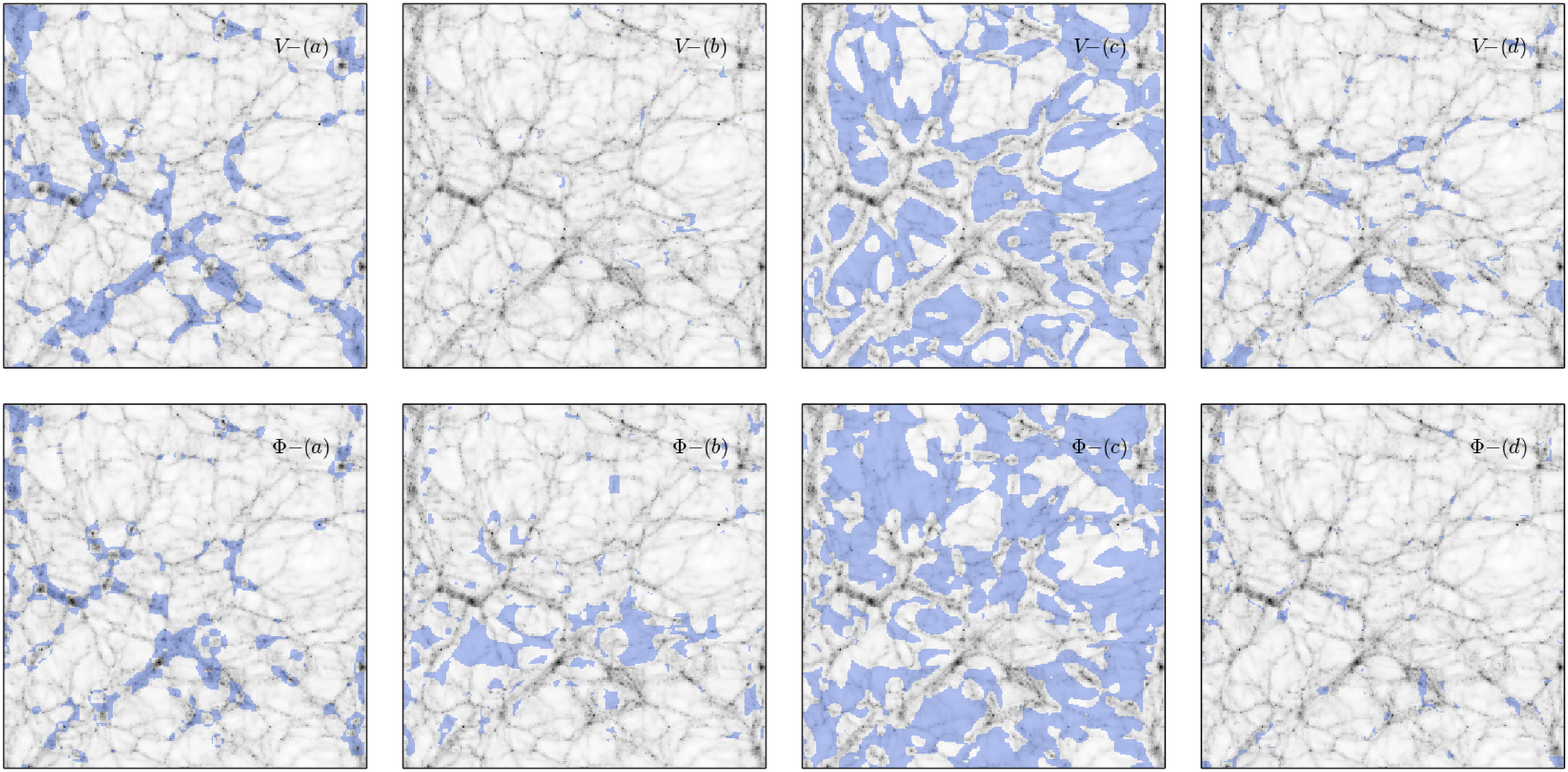}
\end{center}
\caption{ \label{fig:T_V_sim_snapshot_detailed} Detailed morphology comparison between 
kinematic web ({\it upper panels}) and dynamical web ({\it lower panels}), assuming 
threshold $\lambda_{th}=0$. 
From left to right, different panels highlight morphologies types of (a) filaments with 
positive $s^{v/\phi}_1$, (b) filaments with negative $s^{v/\phi}_1$, (c) sheets with 
negative $s^{v/\phi}_1$ and (d) sheets with positive $s^{v/\phi}_1$. 
Although the filamentary structure presented in Figure. (\ref{fig:T_V_sim_snapshot}) are very different, regions shown in the first column are quite similar between these two
algorithms. 
The distortion of the dynamical filaments is due to the contribution from the 
second column, i.e. filaments with negative $s^{(v/\phi)}_1$. 
}
\end{figure*}

\section{Lagrangian Dynamics}
In this section, we will briefly review the Lagrangian evolution of a fluid element in
Newtonian cosmology, especially concentrate on the evolution of the velocity gradient 
tensor $A_{ij}$ and the Hessian matrix of gravitational potential $\Phi_{ij}$. 
Before the shell-crossing, the dynamical equations could be expressed as 
continuity equation, Euler equation as well as Poisson equation \citep{P80,BCGS02},  
\begin{eqnarray}
\label{eqn:Euler_basic_eq}
 \frac{\partial \delta}{\partial \tau} + \nabla \cdot [(1+\delta)\vv] = 0, 
\qquad\qquad\qquad\qquad \nonumber \\
\frac{\partial \vu}{\partial \tau} + \mH(\tau) \vu(\vx, \tau) + 
\vu(\vx, \tau)\cdot \nabla \vu (\vx, \tau) =  -\nabla \Phi(\vx,\tau) \qquad \nonumber\\
\nabla^2 \Phi(\vx, \tau) = 4\pi G \bar{\rho} ~ \delta(\vx, \tau).
\qquad\qquad\qquad\qquad
\end{eqnarray}
As we are interested in the Lagrangian evolution in this paper, the continuity equation
could be further expressed as
\begin{eqnarray}
\label{eqn:Lag_continuity}
 \frac{d\delta}{d \tau} + (1+\delta) \theta  = 0,
\end{eqnarray}
where $d/d\tau$ is Lagrangian total derivative, and $\theta = \nabla \cdot \vu$ 
is the velocity divergence. It is then necessary to consider the Lagrangian equation 
of the velocity gradient tensor $A_{ij}$, which could simply be obtained by taking
the gradient of the second equation in equation (\ref{eqn:Euler_basic_eq})
\begin{eqnarray}
\label{eqn:Aij_evol_formal}
\frac{d A_{ij}}{d \tau} + \mH(\tau) A_{ij} + A_{i}^{~k} A_{kj} = -\Phi_{ij}
\end{eqnarray}
Following the standard treatment, the source term $\Phi_{ij}$ is decomposed as
the trace $\nabla^2 \Phi$ and the traceless part $\varepsilon_{ij}=\Phi_{ij}-
I_{ij}\nabla^2\Phi/3$. While the trace $\nabla^2 \Phi$ simply relates to 
the density via Poisson equation, however, no time evolution equation exists 
for the $\varepsilon_{ij}$ in the Newtonian cosmology. 
However, it is possible to write down the evolution and constraint equations for 
all gravitational fields in general relativity, including the tidal tensor or 
the electric part of Weyl tensor $\varepsilon_{ij}$ \citep{E71,MPS93,BJ94},
\begin{eqnarray}
\label{eqn:eij_LTA}
\frac{d}{d\tau} \varepsilon_{ij} + \mH(\tau)  \varepsilon_{ij} 
+ \theta \varepsilon_{ij}  + I_{ij} \sigma^{kl} 
\varepsilon_{kl} - 3 \sigma^k_{~(i} \varepsilon_{j)k}  \qquad \qquad \nonumber \\
+ \epsilon^{kl}_{~~(i} \varepsilon_{j)k} \omega_l
- \nabla_k  \epsilon^{kl}_{~~(i} \mu_{{j)l}}  =  
 - 4\pi G \rho a^2 \sigma_{ij}, \qquad 
\end{eqnarray}
where $\sigma_{ij}=A_{ij}-\theta I_{ij}/3$ is the traceless velocity shear tensor, 
$\omega_{ij}$ is the antisymmetric vorticity tensor, $\mu_{ij}$ is the magnetic
part of Weyl tensor. 
Parenthesized subscripts denote the symmetrization, and $\epsilon_{ijk}$ 
is the total antisymmetric Levi-Civita tensor. 
Without the terms involving $\mu_{ij}$, the equation is pure local and therefore closed
together with equation (\ref{eqn:Lag_continuity}) and (\ref{eqn:Aij_evol_formal}). 
On the other hand, with the non-vanishing $\mu_{ij}$ term, the dynamics of the tidal 
tensor and velocity gradient become much more complicated \citep{E71,BJ94}. 
Therefore in the following, we will simply assume $\mu_{ij}= 0$. 

Finally, given the evolution of $A_{ij}$ in Eq. (\ref{eqn:Aij_evol_formal}),  the evolution 
equation of invariants $s^{v}_i$ of velocity could be derived straightforwardly \citep{WS14}
\begin{eqnarray}
\label{eqn:inv_dyn_full}
&& \frac{d }{d\tau} s^v_1 + \mH(\tau)s^v_1 -  (s^v_1)^2 +  2 s^v_2  = \Phi_{i}^{~i} 
 \nonumber \\
&& \frac{d}{d\tau} s^v_2  + 2\mH (\tau)s^v_2  - s^v_1 s^v_2  + 3 s^v_3  = s^v_1
\Phi_{i}^{~i}  + \Phi_{ij}A^{ji}   \nonumber \\
&& \frac{d}{d\tau} s^v_3  + 3\mH(\tau) s^v_3 -s^v_1 s^v_3 = s^v_2 \Phi_{i}^{~i} + s^v_1 \Phi_{ij}A^{ji} + \Phi_{i}^{~j}A_{j}^{~k}A_{k}^{~i} \nonumber \\
\end{eqnarray}

\subsection{Zel'dovich Approximation}
In Lagrangian dynamics, the mass element moves in the gravitational field along 
the trajectory 
\begin{eqnarray}
\vx(\vq, \tau)=\vq+\vpsi(\vq, \tau),
\end{eqnarray}
from the initial Lagrangian position $\vq$ to Eulerian coordinate $\vx$, where 
$\vpsi$ is the displacement. 
To the first order, i.e.\ the Zel'dovich approximation, the displacement 
$\vpsi(\vq, \tau)$ is simply given by  \citep{ZA70}
\begin{eqnarray}
\label{eqn:ZA_def}
\nabla_{q} \cdot \Psi(\vq, \tau) = - D(\tau) \delta(\vq),
\end{eqnarray}
where $D(\tau)$ the linear growth factor of density perturbation, and $\nabla_q$ 
denotes the spatial gradient with respect to Lagrangian coordinate.
In Eulerian space, this is equivalent to replacing the Poisson equation with 
\citep{MS94,HB96,BCGS02}
\begin{eqnarray}
\label{eqn:za_assump}
u_i(\vx, \tau) &=& - \frac{2 f(\tau)}{3 \Om(\tau) \mH(\tau) } \nabla_i \Phi(\vx, \tau),
\end{eqnarray}
which then closes the system together with Euler equation (\ref{eqn:Euler_basic_eq}). 
Here, $\Omega_m(\tau)$ is the matter density fraction at epoch $\tau$, and 
$f=d\ln D/ d\ln a$ is the linear growth rate.  
Therefore simply by taking gradient of equation (\ref{eqn:za_assump}), one finds 
$A_{ij}$ and $\Phi_{ij}$ are proportional to each other
\begin{eqnarray}
\label{eqn:aij_phi_za}
A_{ij} &=& - \frac{2 f(\tau)}{3 \Om(\tau) \mH(\tau) } \Phi_{ij},
\end{eqnarray}
Meanwhile since $A_{ij}$ relates to deformation tensor by 
\begin{eqnarray}
A_{ij} =  \mH f [\Psi_{i}^{~k}(I_{kj}+ \Psi_{kj})^{-1}], 
\end{eqnarray}
one could also derive the relation between $\Psi_{ij}$ and $\Phi_{ij}$
\begin{eqnarray}
\Phi_{ij} =  - \frac{3\Omega_m}{2} \mH^2 \Psi_{ij} 
\end{eqnarray}
given $\Psi_{ij}$ is small.

Because of equation (\ref{eqn:aij_phi_za}), we will only concentrate on the evolution
of the velocity invariants $s^{v}$ in the rest of the section. 
As shown in \cite{WS14}, the dynamics of $A_{ij}$ can also be derived starting 
from the simplified Euler equation 
\begin{eqnarray}
\bvu^{\pri} =  \frac{d \bvu}{d D}= 
\left( \frac{\partial}{\partial D} + \bvu\cdot \nabla \right ) \bvu= 0,
\end{eqnarray}
where we have defined the rescaled velocity $\bvu=\vu/D^{(v)}$ where $D^{(v)}(\tau)=dD/d\tau=\mH f D$, and change the time variable $\tau$ into the linear 
growth rate $D$. For the velocity gradient tensor, we similarly define the rescaled quantity
$\bA_{ij}=A_{ij}/D^{(v)}$, and obtain
\begin{eqnarray}
\frac{d \bA_{ij} }{d D} + \bA_{i}^{~k}\bA_{kj} = 0.
\end{eqnarray}
Therefore, the rescaled velocity invariants $\bs^v_1, \bs^v_2, \bs^v_3$ is then defined as
\begin{eqnarray}
\bs^v_i(\tau) = \frac{s^v_i(\tau)}{[D^{(v)}]^i}, \qquad i \in \{1,2,3\}. \quad
\end{eqnarray}
One can then derive a set of ordinary differential equations of reduced invariants:
\begin{eqnarray}
\label{eqn:inv_za_evolution}
(\bs^{v}_1)^{\pri}  - (\bs^v_1)^2 + 2\bs^v_2 &=& 0, \nonumber \\
(\bs^{v}_2)^{\pri} - \bs^v_1 \bs^v_2 + 3\bs^v_3 & =& 0, \nonumber \\
(\bs^{v}_3)^{\pri}  - \bs^v_1 \bs^v_3 & = &0.
\end{eqnarray}
As shown by \cite{W10}, the analytic solution of Eq. (\ref{eqn:inv_za_evolution}) 
could be obtained by taking the third order derivative of $(1/\bs^v_3)$. 
Abbreviating time variable as $D-D_i = d$, the solution is expressed as
\begin{eqnarray}
\label{eqn:za_inv_solution}
\bs^v_1(d) &=& \frac{3\bs^v_3(d_0) d^2 - 2\bs^v_2(d_0)d + \bs^v_1(d_0)}
 { -\bs^v_3(d_0)d^3 + \bs^v_2(d_0) d^2 - \bs^v_1(d_0) d + 1} \nonumber\\
\bs^v_2(d) &=& \frac{-3\bs^v_3(d_0)d + \bs^v_2(d_0)} { -\bs^v_3(d_0) d^3 + 
              \bs^v_2(d_0) d^2 - \bs^v_1(d_0) d + 1} \nonumber \\
\bs^v_3(d) &=& \frac{\bs^v_3(d_0)} { -\bs^v_3(d_0)d^3 + \bs^v_2(d_0) d^2 -
              \bs^v_1(d_0) d + 1}. 
\end{eqnarray}

Therefore in this model, the singularity occurs when the common denominator 
$-\bs^v_3(d_0) d^3 + \bs^v_2(d_0) d^2 - \bs^v_1 (d_0) d + 1 $ becomes zero. 
one also notices that, after rescaling the invariants $\bs^v_2$ and $\bs^v_3$ by 
$|\bs^v_1|^i$, where $i=(2,3)$, these two invariants would approach zero around the 
singularity. 
Finally, since $s^v_3$ never change the sign before singularity, cosmic web 
morphology would remain the same under the assumption that $\lambda_{th}=0$.

\subsection{Nonlinear Local Evolution Model}
\subsubsection{Dynamics}
Assuming irrotational dust model and vanishing magnetic Weyl tensor, equation 
(\ref{eqn:eij_LTA}) is simplified as
\begin{eqnarray}
\label{eqn:epsij_evl_bj94}
	\frac{d}{d\tau} \varepsilon_{ij} + \mH(\tau)  \varepsilon_{ij}
  + \theta \varepsilon_{ij} + I_{ij} \sigma^{kl} 
	\varepsilon_{kl} \qquad \qquad \qquad \nonumber \\
	\qquad \qquad - 3 \sigma^k_{~(i} \varepsilon_{j)k} 
	 =  -4\pi G \rho a^2 \sigma_{ij}. 
\end{eqnarray}
Following \cite{BJ94}, one could conveniently parametrize tensor $A_{ij}$ and 
$\Phi_{ij}$ as
\begin{eqnarray}
\label{eqn:Aij_phi_parametrization}
A_{ij} &=& \frac{1}{3}\theta I_{ij} + \frac{2}{3}\sigma Q_{ij}(\alpha)
  = \frac{1}{3}[ \theta I_{ij} + 2 \sigma Q_{ij}(\alpha)] \nonumber \\
\Phi_{ij} &=& \frac{4\pi}{3} G \bar{\rho} a^2 [ \delta I_{ij} 
   + 2 \varepsilon(1+\delta)Q_{ij}(\beta)  ]
\end{eqnarray}
where $\sigma \le 0$, $\varepsilon \ge 0$ are shear and tidal scalar respectively. 
$\alpha$ and $\beta$ are shear and tides angle, which give the ratios of eigenvalues 
of the shear and tidal tensors. The one-parameter traceless matrix is defined as 
\begin{eqnarray}
\label{eq:Qij_param}
Q_{ij}(\alpha) = {\rm diag} \left [   \cos \left( \frac{\alpha+2\pi}{3}\right),
\cos\left( \frac{\alpha-2\pi}{3}\right), \cos(\frac{\alpha}{3}) \right ]. 
\end{eqnarray}
This definition is uniquely determined by requirements of vanishing trace, 
$Q_{ij}Q^{ji}=3/2$ and $\det[Q_{ij}(\alpha)] = \cos \alpha$. 
Therefore, all possible eigenvalues of traceless
matrix could be characterized by $qQ_{ij}(\alpha)$ with $q \in [0, \infty]$ and 
$\alpha\in [0, \pi]$. As shown in \cite{BJ94}, the matrix also have the following 
property
\begin{eqnarray} 
dQ_{ij}(\alpha) &=& \frac{1}{3} Q_{ij}\left ( \alpha+ \frac{3\pi}{2}\right) d\alpha, 
\nonumber \\
2Q_{i}^{~k}(\alpha)Q_{kj}(\beta) &=& \cos\left( \frac{\alpha-\beta}{3} \right) I_{ij}
+ Q_{ij} (-\alpha-\beta).
\end{eqnarray}

With this parameterization, the Lagrangian equations of motion could be simplify as 
\begin{eqnarray}
\label{eqn:bj94_dynamics_params}
\frac{d\sigma}{d\tau} +\mH\sigma + \frac{1}{3}\sigma(2\theta+\sigma\cos\alpha) = 
 \qquad\qquad \qquad \qquad\quad \nonumber\\
-4\pi G \bar{\rho}a^2 \varepsilon (1+\delta) 
 \cos\left( \frac{\alpha-\beta}{3}\right) \quad \nonumber \\
\frac{d\alpha}{d\tau} - \sigma \sin \alpha = 12 \pi G \bar{\rho} a^2 
\frac{\varepsilon(1+\delta)}{\sigma}  \sin\left(\frac{\alpha-\beta}{3}\right)
\qquad ~ \nonumber\\
\frac{d\varepsilon}{d\tau} -\sigma \varepsilon \cos\left( \frac{\alpha+2\beta}{3}\right) = 
-\sigma \cos\left( \frac{\alpha-\beta}{3}\right) \qquad\qquad \nonumber \\
\frac{d\beta}{d\tau} + 3\sigma \sin\left( \frac{\alpha+2\beta}{3}\right) =
 -\frac{3\sigma}{\varepsilon}\sin\left( \frac{\alpha-\beta}{3}\right).
\qquad \quad 
\end{eqnarray}
Together with the Raychaudhuri equation for $\theta$
\begin{eqnarray} 
\frac{d\theta}{d\tau} + \mH(\tau)\theta + \frac{1}{3}\theta^2 + \frac{2}{3}\sigma^2
= -4\pi G \bar{\rho}a^2 \delta
\end{eqnarray}
and the continuity equation (\ref{eqn:Lag_continuity}), the system is closed. 
Once the solution of physical variables is obtained, one could then derive the 
evolution of velocity invariants $s^v_i$
\begin{eqnarray}
\label{eqn:phyvar_inv_v}
s^v_1 &=& -\theta \nonumber \\
s^v_2 &=& \frac{1}{3} (\theta^2 - \sigma^2 )  \nonumber \\
s^v_3 &=& \frac{1}{27} (-\theta^3 + 3 \theta \sigma^2 - 2 \sigma^3 \cos\alpha)
\end{eqnarray}
as well as  the potential invariants $s^{\phi}_i$ from $\Phi_{ij}$
\begin{eqnarray}
\label{eqn:phyvar_inv_phi}	
s^{\phi}_1 &=&  \delta   \nonumber \\
s^{\phi}_2 &=& \frac{1}{3}\left [ \delta^2 - \varepsilon^2 
               (1+\delta)^2 \right ] \nonumber \\
s^{\phi}_3 &=&  \frac{1}{27} \left [ \delta^3 - 3 \delta
    \varepsilon^2(1+\delta)^2 + 2 \varepsilon^3(1+\delta)^3\cos\beta \right ] 
\end{eqnarray}
Simply by counting the number of dynamical variables, 
one notices that our invariants $\{ s^v_i, s^{\phi}_i\}, ~ i=(1, 2, 3)$
of both $A_{ij}$ and $\Phi_{ij}$ 
fully characterize this nonlinear dynamical model described by physical variables 
$\{ \delta, \theta, \sigma, \varepsilon, \alpha, \beta \}$.

\begin{figure*}
\begin{center}
\includegraphics[width=1.\textwidth]{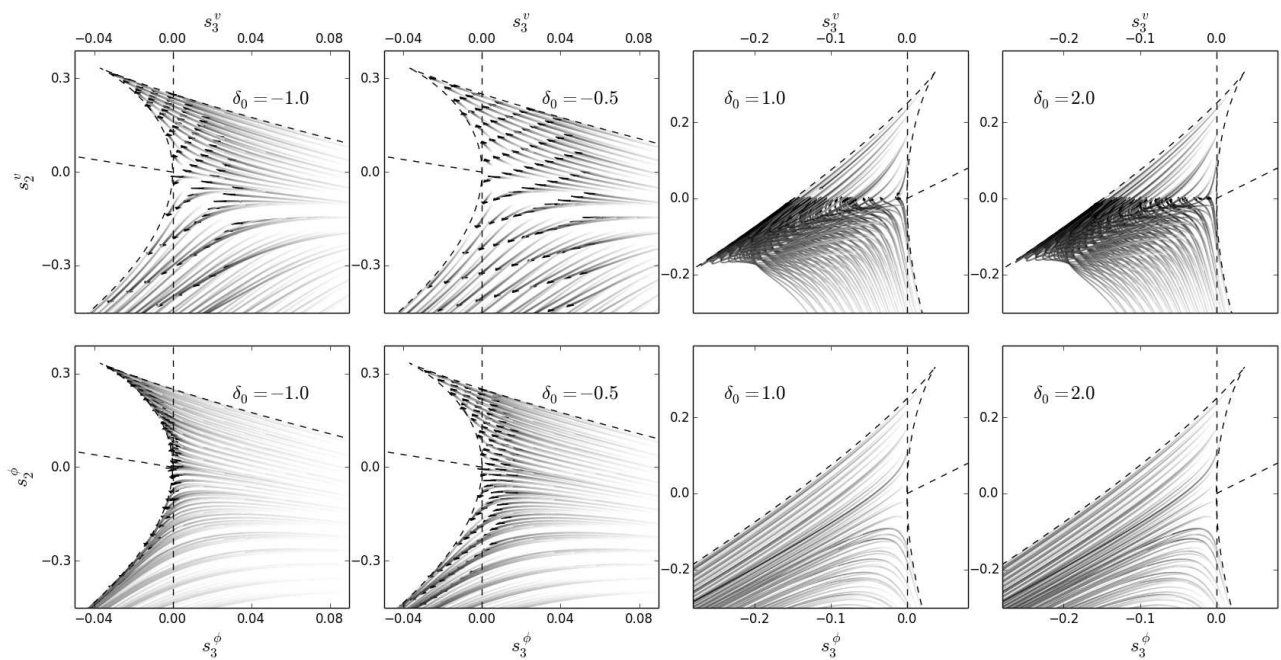}
\end{center}
\caption{ \label{fig:inv_traj_BJ94} 
Comparison between kinematic ({\it upper panels}, defined by tensor $A_{ij}$) and dynamical
 ({\it lower panels}, defined by $\Phi_{ij}$) classification in the invariant space 
for nonlinear local model. 
From left to right, we assume $\delta_0=-1,~ -0.5, ~1$ and $2$. 
The same as Figure (\ref{fig:inv_traj_za_BJ94_comp}), we have normalized 
invariants $s^{(v/\phi)}_i$ such that ${\rm tr}[A_{ij}]={\rm tr}[-\Phi_{ij}] = \pm 1$.  
For underdense perturbation $\delta_0<0$, both tensors exhibit sheet instability, as 
trajectories flow towards voids or filaments, depending the initial conditions. 
However, trajectories in $s^{(\phi)}$ space tends to squeeze towards the boundary 
separating real and complex solutions.
For the overdense perturbation, the singularity of $s^{(v)}$ occurs at $s^{(v)}_2=0$, 
while both $s^{(\phi)}_2$ and $s^{(\phi)}_3$ approach infinity. 
}
\end{figure*}

\subsubsection{Initial Condition}
To specify the initial condition, we first notice that at the linear order
\begin{eqnarray}
\label{eqn:IC_general}
\theta= - \dot{\delta}, \qquad \alpha = \beta, \qquad \sigma_{ij} \propto \epsilon_{ij}. 
\end{eqnarray}
Therefore, among initial values of all six dimensional variable space 
$\{ \delta_0, \theta_0, \sigma_0, \epsilon_0, \alpha_0, \beta_0 \}$, only three of them 
need to be identified initially, either $\{ \delta_0, \epsilon_0, \beta_0 \}$ or 
$\{\theta_0, \sigma_0, \alpha_0  \}$. Particularly, for density and velocity divergence, 
one has
\begin{eqnarray}
\theta_0 = - \frac{d \ln D}{d\tau}(\tau_0) ~ \delta(\tau_0) = -\mH_0 f_0 \delta_0
\end{eqnarray}
For tidal tensor,  since $\alpha_0 = \beta_0$ initially,
to the linear order, one obtains the second-order differential equation of 
$\varepsilon$ the same as the density perturbation $\delta$
\begin{eqnarray}
 \ddot{\epsilon}(\tau) + \mH(\tau) \dot{\epsilon}(\tau) &=& \frac{3}{2} \mH(\tau)
 \Omega_m(\tau) \epsilon(\tau) \nonumber \\
\sigma &=&  - \dot{\epsilon}(\tau). 
\end{eqnarray}
Therefore, $\varepsilon \propto D(\tau)$ at the first order, where $D(\tau)$ is the 
linear density growth rate. And similarly $\sigma_0$ relates to $\epsilon_0$ via 
$	\sigma_0 =  -\mH_0 f_0 \epsilon_0 $.
Moreover, it is also equivalent to specify e.g. the velocity invariants 
$\{ s^v_1, s^v_2, s^v_3 \}$ via
\begin{eqnarray}
\theta (\tau_0) &=& - s^v_1~ (\tau_0) \nonumber\\
\sigma (\tau_0) &=& -\sqrt{(s^v_1)^2  - 3 s^v_2}~ (\tau_0) \nonumber \\
\cos\alpha (\tau_0) &=& \frac{2(s^v_1)^3 - 9 s^v_1 s^v_2 + 27 s^v_3}{2[(s^v_1)^2 -
 3s^v_2]^{3/2}}  ~(\tau_0).
\end{eqnarray}

\subsection{Nonlinear Deformation Tensor}
After solving the dynamical system, one is also able to derive the evolution of deformation
tensor. Before multi-streaming, the displacement of a particle relates to the velocity 
simply by equation $x_i(\vq, \tau)= q_i +\int_{\tau_0}^{\tau}v_i(\tau^{\pri})  d\tau^{\pri} $. 
By taking both the spatial gradient with respect to $q_i$ and the time derivative to this 
equation, one derives the differential equation of the Jacobian matrix $J_{ij}$ 
\begin{eqnarray}
\label{eqn:deform_de}
\frac{d J_{ij}}{d\tau} = A_{i}^{~k} J_{kj}. 
\end{eqnarray}
Assuming the tensor $J_{ij}$ and $A_{ij}$ could be simultaneously diagonalized, 
and denoting their eigenvalues as $\eta_i$ and $\lambda_i$ respectively, 
the solution of above equation could simply be expressed as, 
\begin{eqnarray}
\eta_i (\tau) &=& \eta_i(\tau_0) \exp \left [  \int_{\tau_0}^{\tau} 
\lambda_i(\tau^{\pri})  d\tau^{\pri} \right ] \nonumber \\
 &=& \eta_i(a_0) \exp \left [  \int_{a_0}^{a} 
\frac{\lambda_i(a^{\pri})}{a^{\pri}\mH(a^{\pri})}  da^{\pri} \right ]
\end{eqnarray}
where the initial value $\eta_i (a_0)$ relates to that of $\lambda_i(a_0)$ as
\begin{eqnarray}
\eta_i (a_0)  = \frac{1}{1-\tilde{\lambda}_i(a_0) }, \quad where ~ 
\lambda_i (a_0) = \tilde{\lambda}_i (a_0) \frac{d\ln D}{d\tau} (a_0)
\end{eqnarray}
and $d\ln D/d\tau = \mH f$. One could easily check that above equation holds for 
Zel'dovich approximation.

\begin{figure*}
\begin{center}
\includegraphics[width=1\textwidth]{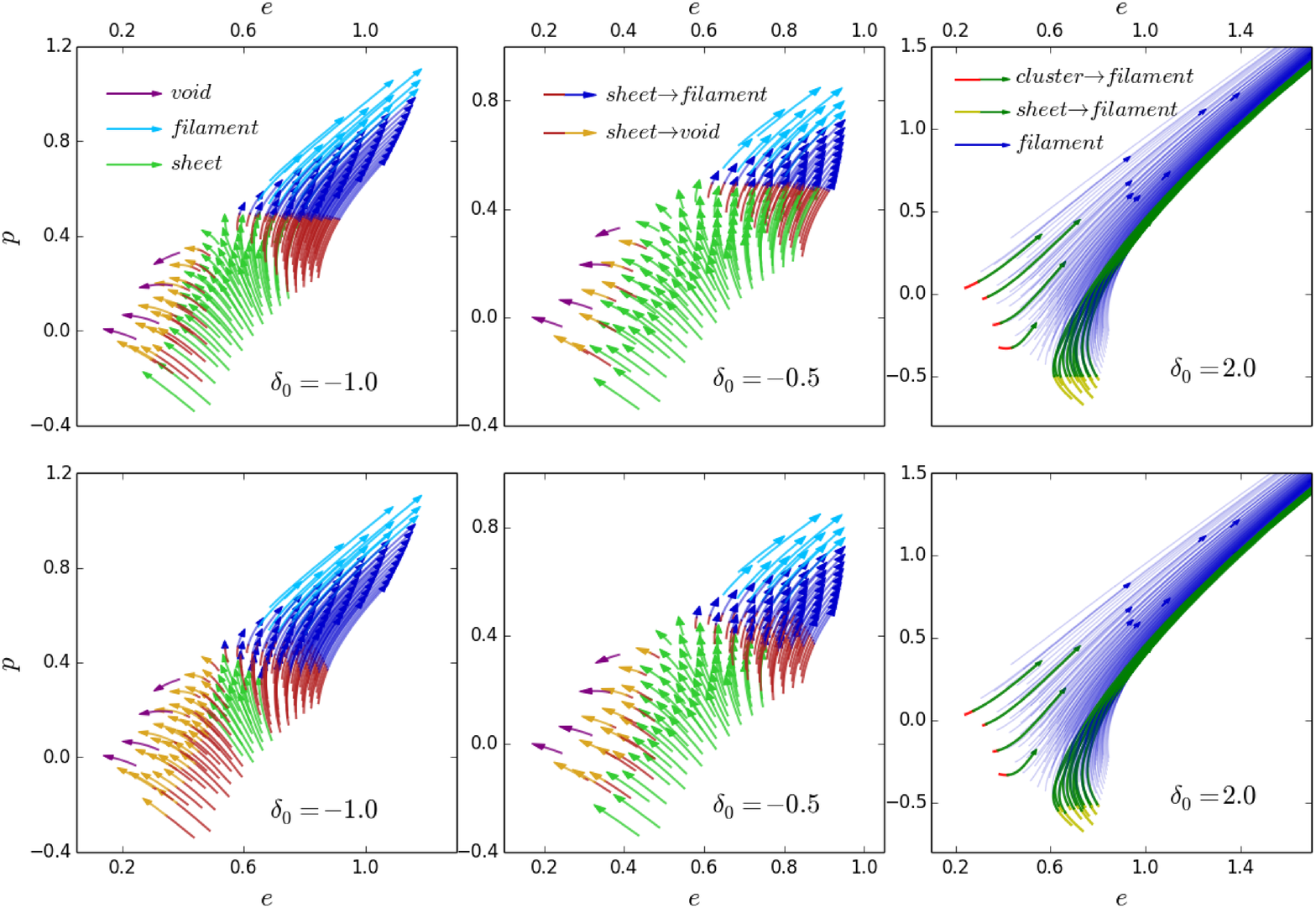}
\end{center}
\caption{ \label{fig:comp_dft_lth0} The evolution of ellipticity $e$ and prolaticity $p$
colored by kinematic  ({\it upper panels}) and dynamical ({\it lower panels}) classification
algorithms, assuming threshold eigenvalue $\lambda_{th}=0$.
All trajectories are calculated via equation (\ref{eqn:deform_de}) with $A_{ij}$ 
supplemented by the nonlinear local model. Morphological evolutions are displayed 
by segmented color arrows, with the discontinuity point indicating the epoch of 
the transition. 
}
\end{figure*}

\section{The Evolution of Cosmic Web}
Given the dynamical equations presented in the last section, one could simply 
integrate the set of ordinary differential equations with appropriate 
initial condition.
In this section, we will present our results for both velocity and potential 
invariants in Zel'dovich approximation as well as the nonlinear local model. 
After comparing these two models in section 4.1, we will mainly concentrate on 
the latter and discuss the differences between the dynamical and kinematic morphology
classifications in section 4.2. In section 4.3, we will then briefly comment on the 
practical freedom of eigenvalue threshold $\lambda_{th}$ and the ambiguity of the 
cosmic web definition.

\subsection{From Zel'dovich Approximation to the Nonlinear Evolution of Cosmic Web}
The pioneering work of the cosmic web evolution by \cite{ZA70} starts with  
the density perturbation as a function of linear growing eigenvalues of 
the deformation tensor $ 1+\delta = 1/ \prod_i[1+ D(\tau) \lambda^{\psi}_i(\tau_0) ] $, 
with $\lambda^{\psi}_i$ being the eigenvalue of tensor $\Psi_{ij}$. 
Despite its simplicity, it suggests that the gravitational collapse would 
generally approach a one-dimensional `plane-parallel' singular solution first
as the probability measure of having two or more same eigenvalues is zero. 
Kinematically, however, since the deformation tensor $ \Psi_{ij} $ grows linearly,
the speed of the collapse is the same for all eigenvalues. 
It means that the morphological type \footnote{assuming eigenvalue threshold 
$\lambda_{th}=0$} inferred from $A_{ij}$ and $\Phi_{ij}$ would always be the 
same before reaching the singularity. 
This statement is valid for both overdense and underdense regions, as 
already seen from the analytical solution (\ref{eqn:za_inv_solution}) of 
$s_i$ in ZA and subsequent discussions thereafter.

In the first row of Figure (\ref{fig:inv_traj_za_BJ94_comp}), we display the 
evolution trajectories in the invariant space for this model. For better presenting the 
result, all invariants are normalized according to equation (\ref{eqn:si_normalize})
with the constant $c=|s_1|$ so that $\tilde{s}_1$ would only take values $1$ or $-1$. 
From left to right, different panels assume various initial conditions 
characterized by the linear density perturbation $\delta_0$ at $a=1$. 
For $\delta_0<0$, we plot all trajectories from initial epoch $a_0 = 10^{-3}$ to 
the present $a=1$; however, for $\delta_0>0$, they will end until the first singularity. 
In the first two panels, all trajectories simply diverge from the origin 
in the $s_2-s_3$ plane without changing categories.
On the other hand, the overdense trajectories approach the first shell-crossing as 
$s^v_2/(s^v_1)^2 \to 0$ and $ s^v_3/(s^v_1)^3 \to 0$. It corresponds to 
a characteristic equation $\lambda^2 (\lambda + 1 ) = 0$, indicating a one-dimensional
collapse with no motion in the other two dimensions.  
Since eigenvalues of $A_{ij}$ are proportional to 
$D(\tau)\xi_i(\tau_0)/[1+ D(\tau) \xi_i(\tau_0) ], ~ i \in \{1,2,3 \}$, 
this occurs when the smallest eigenvalue goes toward $-\infty$ 
while the other two are still finite.

On the other hand, as already noticed by \cite{CPS94} and \cite{BJ94}, 
unlike in ZA, the gravitational collapse in the nonlinear local model
would generally approach filamentary solution, and the sheet structure is usually 
unstable. Although \cite{CPS94} attributed it to the neglect of the
magnetic part of Weyl tensor $\mu_{ij}$, \cite{BJ94} suggested that 
the nonlinear coupling between velocity shear $\sigma_{ij}$ and tidal tensor 
$\varepsilon_{ij}$ in equation (\ref{eqn:epsij_evl_bj94}), i.e.
the term $3\sigma^k_{~(i}\varepsilon_{j)k}-I_{ij} \sigma^{kl}\varepsilon_{kl}$, 
is responsible for this instability. 
Following their arguments, this term with sheet configuration 
$\alpha\approx \beta \approx 0$, has the signature opposite to the sign of 
$\varepsilon_{ij}$ and therefore slows down the growth of tides. Whereas filamentary
configuration with $\alpha\approx \beta \approx \pi$, on the contrary, would
grow due to this term.
This could also be seen from the third equation of (\ref{eqn:bj94_dynamics_params}), 
given that $\cos((\alpha+2\beta)/3)$ approaches unity during the collapse.

In the invariants space, the collapsing filaments mainly correspond to region 
with $s^{(v)}_3 <0$ with condition $s^{(v)}_1>0$. 
From the definition of $s^{(v)}_3$ in equation (\ref{eqn:phyvar_inv_v}), the 
first term $-\theta^3$ is always positive, and the second one $3\theta\sigma^2$ 
is negative. 
For filamentary regions with $ \alpha \approx \pi $, the third term 
$-2\sigma^3\cos\alpha$ is obviously negative. 
However, even when $\alpha$ becomes closer to $0$, as long as the velocity shear 
$\sigma$ grows at a similar speed to $\theta$, 
the term $3\theta\sigma^2$ would dwarf other contributions and therefore form 
filamentary configurations.
Meanwhile, as shown from the last two panels of Figure (\ref{fig:inv_traj_za_BJ94_comp}), 
one notices that the singularity occurs at $s^v_2/|s^v_1|^2 = 0$. 
Since $s^v_2 \propto \theta^2-\sigma^2$, it implies that the velocity divergence 
$\theta$ indeed approaches the infinity at the same speed as velocity shear $\sigma$.

For underdense $\delta_0<0$, \cite{BJ94} found that sufficient large
tides and shear could cause the collapse of some initial expanding perturbations. 
Moreover, from the invariants space of Figure (\ref{fig:inv_traj_za_BJ94_comp}), we could
see a similar morphological instability towards voids or filaments, 
depending on the balance between the initial shear $\sigma$ and divergence $\theta$. 
Since the boundary separating sheets with others is simply $s^{(v)}_3=0$, it 
manifests itself as a universal decay of $s^{(v)}_3$ across the entire two-dimensional 
parameter space. 
For better understanding, we first write down the rescaled invariant as
\begin{eqnarray}
\frac{s^{(v)}_3}{ |s^{(v)}_1 |^3} \propto  -1 + \frac{\sigma^2}{\theta^2} \left 
( 3 - 2 \frac{\sigma}{\theta} \cos \alpha \right) , 
\end{eqnarray}
where $\theta>0$ and $\sigma<0$. Our numerical calculation indicates that this 
ubiquitous decay of $s^{(v)}_3$ actually originates from various contributions 
very differently. 
Although $\sigma$ and $\theta$ all grow as $\dot{D}(\tau)$
at the linear order, the nonlinear evolution of both $\sigma^2/\theta^2$ and 
$ - 2\sigma/\theta \cos\alpha$ then depends on the initial values.
For the most part of the parameter space, the term in the parentheses would decay while 
$\sigma^2/\theta^2$ grows. However, the opposite could also happen for very small 
ratio of $\sigma/\theta$.
Furthermore, the ultimate morphology of this instability after $s^{(v)}_3<0$ would depend 
primarily on the value of $s^{(v)}_2$, and slightly on $s^{(v)}_3$. 
Given the parametrization of $A_{ij} = diag [\lambda_1, \lambda_2, \lambda_3] =
(\theta I_{ij} + 2\sigma Q_{ij}(\alpha) )/3$, 
when $\theta^2 \gg \sigma^2$, even the smallest eigenvalue becomes positive 
regardless of the value of $\alpha$, so the fluid element would evolve to voids. 
On the other hand, if $\theta^2 < \sigma^2$, the smallest and the
medium eigenvalues become negative but not the largest since we assume 
$\sum_i \lambda_i >0$ and $\prod_i \lambda_i > 0$, then it will become filament. 

\begin{figure*}
\begin{center}
\includegraphics[width=1.\textwidth]{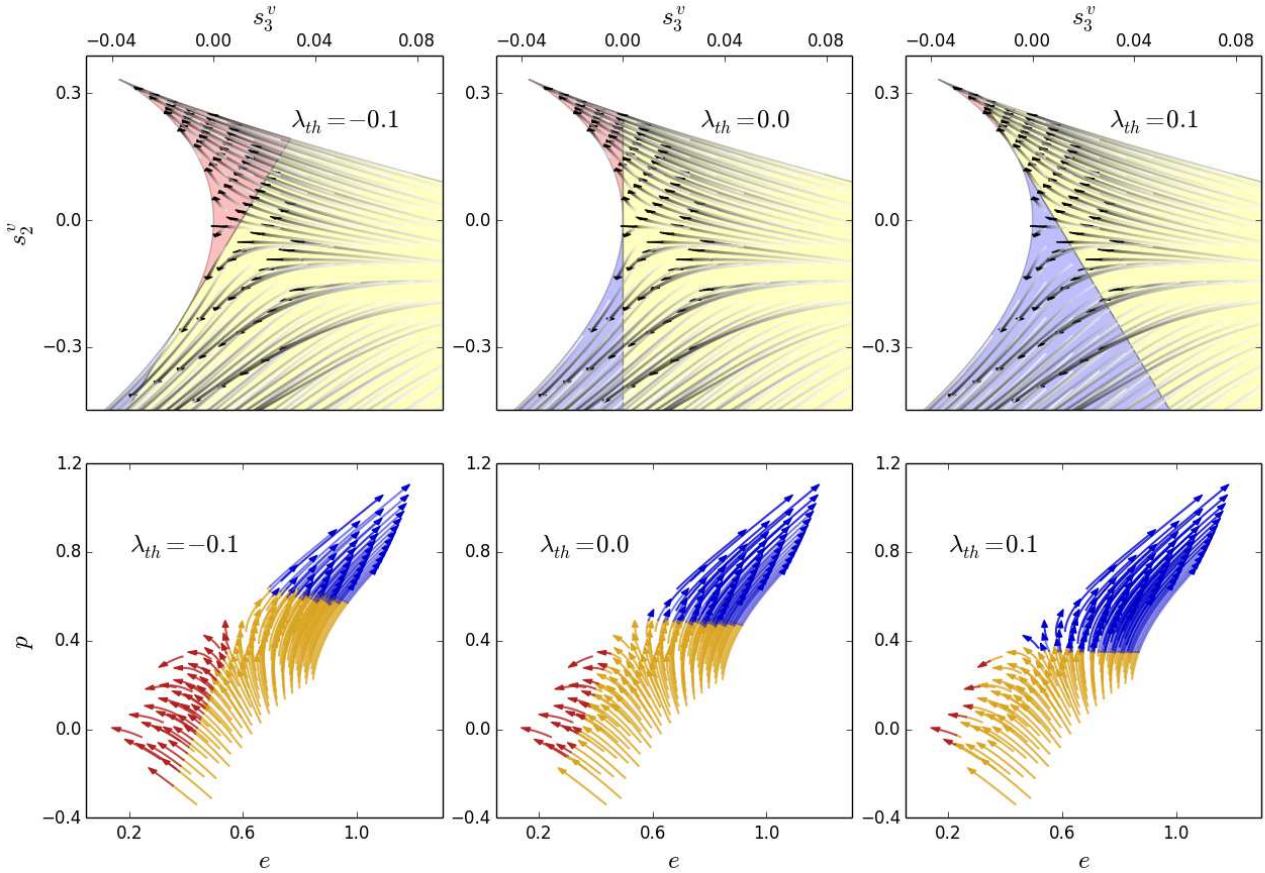}
\end{center}
\caption{ \label{fig:inv_dft_th} The effects of changing eigenvalue threshold 
$\lambda_{th}$ in both velocity invariant space and deformation tensor. 
Here we only present the situation for $\delta_0=-1$. 
Since all trajectories in the upper panels are normalized by $|s^{(v)}_1(a)|^n$, where 
$n=(2, 3)$, this corresponds to a time-dependent threshold $\lambda^{\prime}_{th} (a) = 
\lambda_{th} |s_1(a)| $ with constant $\lambda_{th} = -0.1, 0$ and $0.1$ respectively.
}
\end{figure*}

\subsection{Dynamical and Kinematic Classifications}
The disagreement between the kinematic and dynamical classification algorithms
displayed in Figure. (\ref{fig:T_V_sim_snapshot}) highlights the deviation between
the nonlinear velocity gradient $A_{ij}$ and potential Hessian matrix $\Phi_{ij}$.
Intuitively, one could argue that the anisotropic gravitational forces would affect
the trajectory of a test particle further in time than the velocity gradient, 
and therefore might be a less faithful representation of the current cosmic web. 
Quantitatively, this could be addressed by a direct comparison of the nonlinear 
evolution of these two tensors in, e.g. the nonlinear local model.  
Before proceeding, we would first like to examine Figure. (\ref{fig:T_V_sim_snapshot}) 
in more details. 
To improve the visual impression of dynamical classification algorithm, \cite{FR09} 
suggested to apply a nonzero $\lambda_{th}$ of an order of unity based on the 
argument of spherical collapse model. 
As illustrated in Figure (\ref{fig:inv_space_th}), a negative threshold, of the
tensor $-\Phi_{ij}$ in our convention, would indeed shrink the filamentary region, 
which appears to be responsible for the distorted cosmic web structure, 
and meanwhile increase sheets and decrease clusters.

On the other hand, since the dynamical trajectories differ significantly for 
positive and negative $s^{(v/\phi)}_1$ in the invariant space, it is
convenient to further divide both filament and sheet morphologies based on 
the sign of $s^{(v/\phi)}_1$.
In Figure (\ref{fig:T_V_sim_snapshot_detailed}), we perform this detailed 
comparison for the same simulation snapshot as in Figure (\ref{fig:T_V_sim_snapshot}). 
Interestingly, given almost completely different structures in Figure
(\ref{fig:T_V_sim_snapshot}), filaments with positive $s^{(v/\phi)}_1$, 
shown in the first column of the figure, exhibit a very similar pattern for 
both tensors. Moreover, the major contribution to the dissimilarity come from 
the filaments with negative $s^{(v/\phi)}_1$, as shown in the second column, 
where much more regions are classified as this type for tensor $\Phi_{ij}$ 
than $A_{ij}$.

Assuming the nonlinear local model, we then plot in Figure. (\ref{fig:inv_traj_BJ94}) 
the dynamical evolution of invariants for tensors $A_{ij}$ and $-\Phi_{ij}$ together, 
who are indeed comparable as we have already rescaled the invariants so that 
$\tr[A_{ij}] = \tr[-\Phi_{ij}] = \pm 1 $. 
For underdense perturbation $\delta_0<0$, one sees that morphologies from both 
tensors exhibit similar sheet instability, as all trajectories evolve towards 
the voids or filaments. 
Meanwhile, it is obvious that trajectories in $s^{\phi}$ space squeeze
towards the boundary separating real and complex solutions.  By definition, 
this suggests at least two of eigenvalues should be closer to each other than 
that of tensor $A_{ij}$.
Since $\Phi_{ij}$ is simply proportional to $A_{ij}$ initially, an immediate 
consequence is the enrichment of potential classified filaments with negative 
density perturbation $\delta$, which is exactly what has been observed 
in the simulation.

Physically, at least two factors are responsible for such behavior, as suggested
by the nonlinear local model. 
The first is that the angle $\beta$ approaches $\pi$ from the value of $0$ 
faster than $\alpha$, and therefore it would produce more filamentary structures. 
On the other hand, since the evolution equation of the tidal tensor 
(equation \ref{eqn:epsij_evl_bj94}) is sourced by both density $\rho$ (instead 
of density perturbation $\delta$) and the shear tensor $\sigma_{ij}$, 
$\varepsilon_{ij}$ grows as $\varepsilon (1+\delta)$ compared with $\sigma$ for 
shear tensor (equation \ref{eqn:Aij_phi_parametrization}). 
Consequently, the rescaled tidal tensor  
$\widetilde{\varepsilon}_{ij} \propto \varepsilon (1+\delta) Q_{ij} (\beta)/|\delta| $ 
would grow slower than the rescaled shear tensor $\widetilde{\sigma}_{ij} \propto
\sigma Q_{ij} (\alpha) /|\theta| $ for underdense perturbation $0>\delta>-1$, for in 
general $ |\varepsilon (1+\delta) /\delta| < |\sigma / \theta| $.   
Therefore, the differences between eigenvalues $\Delta \lambda$ are usually narrower
for tensor $\Phi_{ij}$ than $A_{ij}$. 
From the definition of $s^{v/\phi}_2$ in equation (\ref{eqn:phyvar_inv_v}) and 
(\ref{eqn:phyvar_inv_phi}), this 
corresponds to a slower motion of invariant $s^{(\phi)}_2$ than $s^{(v)}_2$, 
as shown in Figure (\ref{fig:inv_traj_BJ94}). 

For overdense perturbation $\delta_0>0$, $s^{\phi}_i$ also evolves very differently 
than $s^{(v)}_i$.
Unlike velocity invariants, where the first singularity occurs at 
$s^{(v)}_2/|s^{(v)}_1|^2 \to 0$, both $s^{(\phi)}_2/|s^{(\phi)}_1|^2 $ and
 $s^{(\phi)}_3/|s^{(\phi)}_1|^3 $ approach infinity as 
$\varepsilon (1+\delta) /\delta \to \infty $,
which again is due to the source term of the evolution equation of tidal field 
$\varepsilon_{ij}$.
However, this does not necessarily suggest the kinematic and dynamical morphologies
differ in this regime.
For vanishing threshold $\lambda_{th}=0$, or even some reasonable nonzero values, 
clusters and sheets identified with both tensors will turn to filaments very soon 
so that no significant differences would emerge.

This could also be seen with the help of the evolution of deformation tensor. 
In Figure.\ (\ref{fig:comp_dft_lth0}), we plot the evolution of ellipticity $e$
and prolaticity $p$, colored by the kinematic morphologies in upper panels and 
dynamical categories in the lower ones. 
Morphological changes are characterized by segmented color arrows, with the 
discontinuity point reflecting the epoch of the morphology transition.
The trajectories are calculated via equation (\ref{eqn:deform_de}) with $A_{ij}$
supplemented by the nonlinear local model.
As expected, the overdense perturbations, shown in the third column, would evolve 
from various initial values towards much higher $e$ and $p$ as they become filaments.
Moreover, both velocity and potential classifications display very similar morphology
categorizations.

For the underdense region, we show both $\delta_0=-1$ and $\delta_0=-0.5$ in the first
two columns. Consistent with the bifurcate evolution in the invariant space, trajectories 
with various initial ellipticity and prolaticity flow towards the opposite directions 
in $e-p$ plane. One is the spherical void region with $e\sim 0$ and $p\sim 0$, and 
the other is the non-spherical prolate filament region with  $e \approx p \sim 1 $, 
while the non-spherical sheets reside in between. 
Since both upper and lower panels display the same geometrical evolution of a fluid
element, the color scheme shows that gravitational potential-based algorithm in
general would identify more filaments and fewer sheets than kinematical classification.

\subsection{Eigenvalue Threshold and the Ambiguity of Cosmic Web Definition}

Practically, a nonzero threshold is usually applied in the algorithm to `optimize' 
the visual impression of the cosmic web.
As already shown in Figure (\ref{fig:inv_space_th}),  a negative $\lambda_{th}$ 
would indeed help to reduce the otherwise excessive 
filaments and sheets, meanwhile increase the volume fraction of void regions.
However, geometric deformation of a fluid element is well defined
by quantities like ellipticity and prolaticity. 
In Figure. (\ref{fig:inv_dft_th}), we highlight the morphologies variations in both 
velocity invariant space and the deformation $e-p$ plane.
Since all trajectories in the upper panels are normalized by $|s^{(v)}_1(a)|^n$, where 
$n=(2, 3)$, this corresponds to a time-dependent threshold $\lambda^{\prime}_{th} (a) = 
\lambda_{th} |s_1(a)| $ with constant $\lambda_{th} = -0.1, 0$ and $0.1$ respectively. 
Therefore, the morphology of a fluid element with given shape measurement  
depends on the threshold $\lambda_{th}$, which reflects the ambiguous definition of 
the cosmic web.

This then leads to the question about the purpose of the morphological 
classification and its associated `best' algorithm.
An outstanding visual impression would require both density threshold 
and anisotropic information, like $A_{ij}$ or $\Phi_{ij}$, at various scales. 
For many studies, e.g. the environmental dependence of halo formation, 
it is probably more important to describe quantitatively the entanglement of 
relevant quantities than satisfying the preference of the human brain. 
In this sense, the `morphology classification' is only a simplification 
to the more complicated problem. 
Without any arbitrary tuning of the threshold, the tensor $A_{ij}$ and 
$\Phi_{ij}$ themselves and corresponding rotational invariants are attractive 
quantities as they characterize the underlying physical processes.
Therefore, besides developing various classification algorithms, more efforts 
should be made to understand the detailed evolution of these quantities.

\section{Conclusion and Discussion}

In this paper, we revisited the Lagrangian evolution of various tensors, 
including the velocity gradient tensor $A_{ij}$, the 
Hessian matrix of gravitational potential $\Phi_{ij}$ and the deformation tensor 
$\Psi_{ij}$, for their useful applications in the cosmic web classification.
Unlike previous studies, we performed the investigation in the invariant space, 
defined as coefficients of the characteristic equation of $A_{ij}$ and $\Phi_{ij}$. 
Compared with the eigenvalue representation, this parameter space is much more 
convenient in tracking the dynamical evolution of these tensors. 
We then presented the solution for both Zel'dovich approximation and the nonlinear 
local model. Although the latter model is neither Newtonian nor fully general
relativistic, it is reasonable to assume to be a suitable approximation for our
purpose.

Since one could easily write down the analytical solution of invariants evolution 
in ZA, we reconfirm the fact that cosmic morphologies would not change before 
approaching a one-dimensional singularity in this model. 
However, the nonlinear local model would in general lead to the morphology 
instability and changes. For overdense perturbation, the sheet configurations 
usually collapse to filaments very quickly due to the coupling between tidal
tensor $\varepsilon_{ij}$ and velocity shear $\sigma_{ij}$.  
For underdense regions, however, the sheet could either evolve to void
or filament depending on the balance between the shear $\sigma$ and divergence 
$\theta$ for $A_{ij}$, or the tides $\varepsilon$ and density $\delta$ for 
$\Phi_{ij}$.

Interestingly, our comparison of the invariants trajectories between tensor 
$A_{ij}$ and $\Phi_{ij}$ suggests that different evolving speed of the instability 
is responsible for some distinctions of the cosmic web classified using these two 
tensors. Since both tensors start from the same morphologies initially, 
the squeezed trajectories of $\Phi_{ij}$ in Figure (\ref{fig:inv_traj_BJ94}) 
suggests more abundant filaments with negative $s^{(\phi)}_1$, which is 
exactly what has been observed in the simulation. 
Physically, this is caused by both different evolving speed of tensor angle
$\alpha$ and $\beta$, and the source term of the tidal field evolution 
equation (\ref{eqn:epsij_evl_bj94}).

However, there're some limitations of our approach as well. First of all, the 
dynamics only work before the singularity, therefore very little conclusion 
would be able to make for overdense regions.
Fortunately, for our purpose, it is equally, if not more, important to 
study the underdense perturbations. 
Secondly, since it's still possible that this nonlinear local model would not
fully capture the real dynamics, one need to be cautious about the direct 
comparison between simulation and the model calculation. 
For example, from Figure (\ref{fig:inv_traj_BJ94}), one might also expect 
to observe more potential classified voids than the other algorithm. 
However, the simulation measurement produces somewhat similar fractions of 
voids for these two methods. 
In addition, if one tries to measure the eigenvalue differences 
$\Delta \lambda = |\lambda_i - \lambda_j|, ~i\ne j$ from the simulation, 
the inequality 
\begin{eqnarray}
\label{eqn:deltav_l_deltap}
\Delta \lambda^{(v)} > \Delta \lambda^{(\phi)}
\end{eqnarray}
would only hold for the differences between the largest eigenvalue 
and the other two. 
For the difference between the smallest two eigenvalues, however, both 
tensors have similar distributions, with only slightly asymmetry favoring 
equation (\ref{eqn:deltav_l_deltap}). 
On the other hand, theoretical calculation shows its validity for all three 
$\Delta \lambda$s. 
Nevertheless, whether or not this suggests the failure of the model is 
not clear.

\section*{acknowledgments}
We thank Michael Wilczek and Mark Neyrinck for useful discussions. 
This work has been supported by the Gordon an Betty Moore and Alfred P. Sloan 
Foundations in Data Intensive science.

\appendix
\newcommand{\appsection}[1]{\let\oldthesection\thesection
\renewcommand{\thesection}{\oldthesection}
\section{#1}\let\thesection\oldthesection}

\label{lastpage}

\end{document}